\documentstyle[version2,preprint,aps]{revtex}	
\begin{document}
%
%
%
\preprint{DOE-ER40757-024,CPP-93-24}
\preprint{FSU-HEP-930808}
\preprint{August 1993}
\begin{title}
{\bf PRODUCTION OF $Z$ BOSON PAIRS \\
AT PHOTON LINEAR COLLIDERS }
\end{title}
\bigskip
\author{Duane A. Dicus}
\begin{instit}
Center for Particle Physics, University of Texas at Austin,
Austin, TX 78712, USA
\end{instit}
\begin{center}
and
\end{center}
\author{Chung Kao}
\begin{instit}
Department of Physics, B-159, Florida State University,
Tallahassee, FL 32306, USA
\end{instit}
\begin{abstract}

The $ZZ$ pair production rate in high energy $\gamma \gamma$ collisions
is evaluated with photons from laser backscattering.
We find that searching for the Standard Model Higgs boson with a mass up to,
or slightly larger than, 400 GeV via the $ZZ$ final state is possible
via photon fusion with backscattered laser photons
at a linear $e^+e^-$ collider with energies in the range
600 GeV $< \sqrt{s_{e^+e^-}} <$ 1000 GeV.
\end{abstract}
%
%
%
\newpage
\narrowtext
\section{Introduction}

In many cases, interesting physics processes can be studied with high precision
at linear $e^+e^-$ colliders where the background is usually low and the signal
is much cleaner than that of hadron colliders.
The Next Linear Collider (NLC) is a projected linear $e^+e^-$ collider
with a center of mass (CM) energy ($\sqrt{s_{e^+e^-}}$) of 500 GeV and
a yearly integrated luminosity of about 10 $fb^{-1}$.
In $e^+e^-$ collisions, the Higgs boson of the Standard Model (SM) with a mass,
$M_H$, up to 350 GeV \cite{Barger,Jack} will be observable at the NLC.
Improvements in the technology of laser backscattering have made it
likely that the NLC could be run as a high energy photon collider
\cite{Ginzburg1}-\cite{BBC}.
Photon fusion can become a promising source
to produce and study the Higgs bosons \cite{Higgs}-\cite{Cheung}
of the SM and its extensions when the high energy $\gamma\gamma$ luminosity
at linear $e^+e^-$ colliders is greatly enhanced by laser backscattering.
However, it was recently found \cite{Jikia,Berger},
that the transverse $Z_TZ_T$ pair produced from photon fusion can become
a serious irreducible background and make the Higgs search
via the $ZZ$ decay mode in $\gamma\gamma$ collisions
impossible at the NLC and higher energy linear $e^+e^-$ colliders
if $M_H$ is larger than about 350 GeV.

In this letter, the complete SM calculation of $\gamma \gamma \to ZZ$
is evaluated independently.
A non-linear gauge is used to greatly reduce the number of diagrams
and simplify the Feynman rules.
The total cross section and invariant mass distribution
of $ZZ$ pair at photon colliders is presented and the search for
the SM Higgs boson is examined.
Our cross sections of $\gamma \gamma \to ZZ$
for monochromatic photon photon collisions agree with that
of Ref. \cite{Jikia} where a different non-linear gauge was used,
and Ref. \cite{Berger} where a linear gauge was adopted and unpolarized
initial $e^+e^-$ and laser beams were considered.
We have also checked the total cross section and invariant mass distribution
for $ZZ$ pair production with the polarizations of initial $e^+e^-$
and laser beams as well as CM energies of $e^+e^-$ considered
in Ref. \cite{Jikia},  and have found good agreement.
In addition, we have considered other CM energies of $e^+e^-$ and
other polarizations of the electron positron and laser beams.
Our conclusion as to a viability of a Higgs search
with a realistic energy spectrum for backscattered photons
is slightly more optimistic than that of Ref. \cite{Jikia}
or \cite{Berger}.

\section{Non-Linear Gauge Fixing and Loop Integration}

In the SM, the lowest order $\gamma\gamma ZZ$ coupling comes from
the 1-loop diagrams of the leptons ($l$), the quarks ($q$),
and the physical $W$ boson ($W^{\pm}$) in the unitary gauge.
The Higgs boson has a significant effect on the $W$ loop and
the top quark loop contributions.
The Nambu-Goldstone boson ($G^{\pm}$)
and the Fadeev-Popov ghosts ($\theta^{\pm},\bar{\theta}^{\pm}$)
play important roles in a general gauge and make $W$ loop calculation
unnecessarily complicated.
It has been demonstrated for processes with photons that
a carefully chosen non-linear gauge
\cite{Fujikawa}-\cite{Boudjema}
can remove the mixed vertices of photon-W-G ($A^\mu W^{\pm}_\mu G^{\mp}$)
and Higgs-photon-W-G ($H A^\mu W^{\pm}_\mu G^{\mp}$),
reduce the number of loop diagrams and simplify the Feynman rules.

In this letter, a non-linear $R_\xi$ gauge is introduced to remove
not only the mixed vertices $\gamma WG$ and $\gamma HWG$
but also the vertices $ZWG$ and $ZHWG$.
The gauge fixing terms are chosen to be
\begin{eqnarray}
{\cal L_{GF}} & = & -\frac{1}{\xi_W} f^+f^-
                    -\frac{1}{2\xi_Z} (f^Z)^2
                    -\frac{1}{2\xi_A} (f^A)^2 \\
f^+ & = & \partial^\mu W^+_\mu -i\xi_W M_W G^+
          +i g^\prime B^\mu W^+_\mu \nonumber \\
    & = & \partial^\mu W^+_\mu -i\xi_W M_W G^+
          +i g (-\frac{\sin^2 \theta_W}{\cos \theta_W} Z^\mu
                + \sin \theta_W A^\mu ) W^+_\mu \\
f^Z & = & \partial^\mu Z_\mu -\xi_Z M_Z G^0 \\
f^A & = & \partial^\mu A_\mu
\end{eqnarray}
where $f^-$ is the Hermitian conjugate of $f^+$,
$M_W$ and $M_Z$ are masses of the $W$ and $Z$ bosons,
$g = e/\sin \theta_W$ and $\theta_W$ is the Weinberg angle.
The ghost couplings that depend on the gauge fixing term (1);
and all modified Feynman rules are given in an appendix.
The gauge parameters are all taken to be unity,
$\xi_W = \xi_Z = \xi_A =1$,
which corresponds to a non-linear 't Hooft-Feynman gauge.
In this new gauge, there are 3 pure classes of diagrams for
the $W$ boson ($W$-loop), the Nambu-Goldstone boson ($G$-loop)
and the Fadeev-Popov ghosts ($\theta$-loop)
with the same mass $M_W = M_G =M_\theta$.
Further, the ghost loops contribute -2 times the Nambu-Goldstone boson loops
except for those loops with a $ZZ\theta\theta$ coupling.
In addition to the box (4-point) and the triangle (3-point) diagrams
which appear in the fermion loops, there are also bubble (2-point) diagrams
in the $W$, $G$ and $\theta$ loops:
24 box, 48 triangle and 12 bubble diagrams without the Higgs boson;
8 triangle and 4 bubble diagrams with the Higgs boson; which add up to
96 diagrams in this gauge.
In the linear $R_\xi$ gauge \cite{Berger}, there are 188 diagrams:
108 box, 48 triangle and 6 bubble diagrams without the Higgs boson;
20 triangle and 6 bubble diagrams with the Higgs boson.
The fermion loops are obtained from an earlier calculation of
$gg \to ZZ$ \cite{ggZZ,Glover} with a modification of couplings.
All loop integrations have been calculated with the computer
program LOOP \cite{LOOP,Tini}, which evaluates one loop integrals
analytically and generates numerical data.
The resulting numerical program is checked by replacing the polarization
vector for one of the photons with its four-momentum.
Gauge invariance requires that this yield a vanishing result
which checks all integrals and algebra involved.

\section{Monocromatic $\gamma\gamma$ Collisions}

The amplitude of $\gamma\gamma \to ZZ$ can be written as

\begin{equation}
M_{\lambda_1\lambda_2\lambda_3\lambda_4}
= \epsilon^\mu_1 \epsilon^\nu_2 \epsilon^\rho_3 \epsilon^\sigma_4
  T_{\mu\nu\rho\sigma}(p_1,p_2,p_3,p_4)
\end{equation}
where $\lambda_{1,2}$ and $\lambda_{3,4}$ are the helicities of
the photons and the $Z$'s, the $p$'s are the momenta and
the $\epsilon$'s are the polarization vectors.

The cross sections of $\gamma\gamma \to ZZ$
in different helicity states of $ZZ$ are presented in Figure 1 as a function
of $\sqrt{s_{\gamma\gamma}}$ for both polarizations, $++$ and $+-$,
of the photons.
The parameters used are $\alpha = 1/128$, $\sin^2 \theta_W = 0.230$,
$M_Z = 91.17$ GeV and $M_W = M_Z \cos \theta_W$.
The Higgs mass ($M_H$) is taken to be 300, 400, 500, and 800 GeV.
If not mentioned, the top quark mass ($m_t$) is considered to be 140 GeV.
Also shown is the $++LL$ cross section without the Higgs boson,
which is the same as taking $M_H = \infty$.
As can be easily seen, the $Z_TZ_T$ cross section dominates and almost
approaches a constant as $\sqrt{s_{\gamma\gamma}} > 1$ TeV,
except for $M_H < 300$ GeV where the $++LL$ cross section is larger.
Not shown are the individual contributions from the $W$ loop and fermion loops.
The $W$ loop is usually at least about 10 times larger.
Only in the $++LL$ state and for large $m_t$ and high energy,
can the top quark loop be comparable to the $W$ loop;
and only in the $++LT$ state at low energy, can the fermion loop dominate.
For $M_H = \sqrt{s_{\gamma\gamma}} > 450$ GeV, the  $++TT$ cross section
is almost an order of magnitude larger than that of $++LL$,
which makes the Higgs search in the $ZZ$ mode via photon fusion impossible for
$M_H > 450$ GeV, unless the transverse and longitudinal polarizations
of the $Z$ boson can be distinguished.
All our numerical data agree with those in Ref. \cite{Jikia},
except the cross section for $++LL$ cross section with $M_H = \infty$.

The $m_t$ dependence and the interference between the $W$ loop and fermion loop
for the $++LL$ helicity states are shown in Figure 2 for
$m_t = 120, 160$ and 200 GeV.
The $W$ loop cross section is not sensitive to the top quark mass;
it depends on $m_t$ only in the Higgs width and
therefore is evaluated with $m_t = 160$ GeV only.
The total cross section at $\sqrt{s_{\gamma\gamma}} = M_H = 300-800$ GeV
are also presented in Table I for $m_t = $120, 140, 160, 180 and 200 GeV
where the precise value of $m_t$ is used everywhere.
Several interesting aspects can be learned from Fig. 2 and Table I:
(1) The $W$ loop and the fermion loop interfere destructively.
(2) For $M_H$ below 300 GeV, the total $++LL$ cross section grows with $m_t$,
while for $M_H$ above 700 GeV it decreases as $m_t$ becomes larger.
(3) For $M_H = 400, 500$ and 600 GeV there is a minimum which appears at about
$m_t = 130, 160$ and 180 GeV, respectively.
The $++LL$ cross section always depends on the $m_t$ which appears in the
Yukawa coupling of the top quark to the Higgs boson.
Not shown is the $TT$ cross section which become insensitive to $m_t$ for
$\sqrt{s_{\gamma\gamma}} >$ 500 GeV.

The Higgs boson contributes only to the states with the same photon
helicities and the same $Z$ helicities.
In order to improve the ratio of signal to background while saving most of the
$LL$ signal, we consider a cut on the CM scattering angle
$|\cos(\theta^*)| = |z| < 0.8$ which reduces about $30\%$ of the $++TT$,
and more than $45\%$ of the $+-TT$ background
while saves about $80\%$ of the $++LL$ signal.
For the total cross section, the efficiency of this angular cut and one
with $|z| < \cos( 30^o )$ are presented in Table II
for $\sqrt{s_{\gamma\gamma}} = M_H = 300, 400$ and  500 GeV.

\section{Backscattered Laser $\gamma\gamma$ Collisions}

It has been shown that $\gamma\gamma ZZ$ can hardly be observed
with the Weizs\"{a}cker-Williams photons \cite{Jikia},
because the $\gamma\gamma$ luminosity falls rapidly as the
$\gamma\gamma$ invariant mass increases.
Fortunately, Compton laser backscattering can produce high energy
photons with high luminosity.
The total cross section of $ZZ$ pair production at linear $e^+e^-$ colliders
with backscattered laser photons
is evaluated from the differential cross section of
the photon fusion subprocess $\gamma\gamma \rightarrow ZZ$
with the convolution of photon spectrum.
\begin{eqnarray}
d\sigma_{\lambda_3\lambda_4} & = &
\kappa \int_{4{m_Z}^2/s}^{{y_m}^2} d\tau \frac{dL_{\gamma\gamma}}{d\tau}
[ \frac{1+<\xi_1\xi_2>}{2}d\hat{\sigma}_{++\lambda_3\lambda_4}
 +\frac{1-<\xi_1\xi_2>}{2}d\hat{\sigma}_{+-\lambda_3\lambda_4}], \\
\frac{dL_{\gamma\gamma}}{d\tau} & = &
\int_{\tau/y_m}^{y_m} \frac{dy}{y}
f_{\gamma/e}(y,x) f_{\gamma/e}(\tau/y,x), \\
r & = & M_{ZZ}/\sqrt{s}, \\
\tau & = & \hat{s}/s = r^2, \\
y & = & E_\gamma /E_e, \\
y_m & = & \frac{x}{x+1}, \\
x & = & 4E_e \omega_0/m_e^2
\end{eqnarray}
where
$f_{\gamma/e} =$ photon energy distribution function,
$M_{ZZ} =$ the invariant mass of the $ZZ$ pair,
$E_e =$ the initial electron energy,
$E_\gamma =$ backscattered photon energy,
$\omega_0 =$ the laser photon energy,
$\kappa =$ number of high energy photons per one electron, and
$\xi_{1,2} =$ the mean helicities of the photon beams.
The maximal energy available in the CM frame of $\gamma\gamma$ is
$E_{MAX} = y_m \sqrt{s_{e^+e^-}}$.
We have taken $\kappa = 1$, and $x = 4.8$ which gives $y_m = 0.83$.
As noted in Ref. \cite{Telnov}, if $x > 4.8$,
number of high energy photons will be reduced
by unwanted $e^+e^-$ pair production.
The $f_{\gamma/e}$ and $\xi_i$ are taken from equations (4), (12) and (17)
of Ref. \cite{Ginzburg2}.

The energy spectrum of photons from Compton laser backscattering
depends on the product $2\lambda_e \lambda_\gamma$ \cite{Ginzburg2},
where $\lambda_e =$ the degree of polarization (mean helicity) of the initial
electron (positron) and
$\lambda_\gamma =$ the degree of circular polarization or mean helicity
of the laser beam.
The number of high energy photons increases while the number of
soft photons decreases when $-2\lambda_e \lambda_\gamma$ becomes larger.
We have studied the photon energy spectrum with $x = 4.8$
for three combinations of polarizations of the initial $e^+e^-$ and laser
beams:
(a) $\lambda_e = 0.5$ and $\lambda_\gamma = -1.0$,
polarized $e^+e^-$ and laser beams with $2\lambda_e \lambda_\gamma =-1$ ;
(b) $\lambda_e = 0.5$ and $\lambda_\gamma = 1.0$,
polarized $e^+e^-$ and laser beams with $2\lambda_e \lambda_\gamma =+1$ ;
and (c) $\lambda_e = 0 $ and $\lambda_\gamma = 0$,
unpolarized $e^+e^-$ and laser beams with $2\lambda_e \lambda_\gamma = 0$.
Several interesting aspects have been found:
(1) All of them produce about the same number of photons at
an energy fraction $y_0 = E_\gamma/E_e = 0.7$.
(2) Below $y_0$, the photon luminosity of case (b) is slightly larger than the
others. However, it falls off rapidly for $y > y_0$. Case (a) rises sharply
for $y > y_0$, but yields the smallest number of photons below $y_0$.
(3) In case (c), $2\lambda_e \lambda_\gamma = 0$, the spectrum is almost
flat below $y_0$, and the photon luminosity rises significantly as $y > y_0$.
(4) In $\gamma\gamma$ collisions, the energy fractions $y_1$ and $y_2$
are related by $y_1 y_2 = \tau$. With
$\lambda_{e_1} = \lambda_{e_2} = \lambda_e$
and $\lambda_{\gamma_1} = \lambda_{\gamma_2} = \lambda_\gamma$,
case (a) has the highest photon photon luminosity for
$r = M_{ZZ}/\sqrt{s_{e^+e^-}} > 0.7$ while case (b) dominates for $r < 0.6$.
Case (c) produces a larger number of photons in a broad range of energies.

Our main purpose is to enhance the Higgs signal as much as possible.
The Stokes parameters $<\xi_1\xi_2>$ in Eq. (6) play important roles
in enhancing or reducing the Higgs signal.
To study the effect of $<\xi_1\xi_2>$ with
$\lambda_{e_1} = \lambda_{e_2} = \lambda_e$
and $\lambda_{\gamma_1} = \lambda_{\gamma_2} = \lambda_\gamma$
in photon photon collisions, we have considered two more cases,
(d) $\lambda_e = 0.5$ and $\lambda_\gamma = 0$,
polarized $e^+e^-$ and unpolarized laser beams; and
(e) $\lambda_e = 0$ and $\lambda_\gamma = 1.0$,
unpolarized $e^+e^-$ and polarized laser beams;
in addition to the three cases just considered for the photon energy spectrum.
Since the Higgs appears only in the same helicity states of $\gamma\gamma ZZ$,
we would like to enhance the cross section
of $\hat{\sigma}_{++\lambda_3\lambda_4}$
while reducing $\hat{\sigma}_{+-\lambda_3\lambda_4}$.
We have found that, in case (d), $<\xi_1\xi_2>$ is always positive,
and is enhanced as $M_{ZZ}$ becomes larger.
Case (b) usually has positive $<\xi_1\xi_2>$ and it is the largest
at low $M_{ZZ}$, however it drops rapidly for $r > 0.7$.
In case (a), $<\xi_1\xi_2>$ is usually positive for $r < 0.30$ and $r > 0.63$,
but usually negative in between.
In Case (c), $<\xi_1\xi_2> = 0$.
In case (e), $<\xi_1\xi_2>$ is usually positive for $r < 0.54$ and
$r > 0.76$, but becomes negative in between.
The combination of polarizations $-\lambda_e$ and $-\lambda_{\gamma}$
has the same product $2\lambda_e \lambda_{\gamma}$ as that of
$\lambda_e$ and $\lambda_{\gamma}$, therefore produces the same energy spectrum
but it yields $<\xi_1\xi_2>$ with an opposite sign.
As a combined effect from energy spectrum and the Stokes parameters,
case (d) seems to be the best choice for a the Higgs search
over a broad range of $M_H$, case (a) is the best for
$M_H > 0.7 \sqrt{s_{e^+e^-}}$,
and case (b) is the best for $M_H < 0.6 \sqrt{s_{e^+e^-}}$.

The total cross section of $\gamma\gamma \to ZZ$ in high energy
photon photon collisions with backscattered laser photons is presented
as a function of $\sqrt{s_{e^+e^-}}$ in Table III, for
$m_t = 140$ GeV and $m_H = 300, 400$ GeV and $\infty$ (the background)
and the five combinations of polarizations for the initial $e^+e^-$
and laser beams used for studying the Stokes parameters.
{}From Table III, we can find that
(1) For $M_H$ close to $E_{MAX}$,
$ \lambda_{e_1} = \lambda_{e_2} = 0.5 $ and
$ \lambda_{\gamma_1} = \lambda_{\gamma_2} = -1.0$
produces the largest cross section;
(2) For $M_H$ much smaller than $E_{MAX}$,
$ \lambda_{e_1} = \lambda_{e_2} = 0.5 $ and
$ \lambda_{\gamma_1} = \lambda_{\gamma_2} = 1.0$
produces the largest cross section;
and (3) the unpolarized initial $e^+e^-$ or laser beams,
yield a clear Higgs signal for a broad range of energy.

To study the observability of the Higgs signal as a pronounced peak in
the $ZZ$ invariant mass distribution, we consider the total contribution
without the Higgs boson as the background and show the Higgs signal
with the background in Figures 3 and 4.
The invariant mass distribution of $ZZ$ for $\gamma\gamma \to ZZ$
at the NLC, $\sqrt{ s_{e^+e^-} } = 500$ GeV, is shown in Figure 3
for the three most promising polarizations of initial electron(positron)
and laser beams:
 (a) $ \lambda_{e_1} = \lambda_{e_2} = 0.45 $ and
     $ \lambda_{\gamma_1} = \lambda_{\gamma_2} = -1.0$,
 (b) $ \lambda_{e_1} = \lambda_{e_2} = 0 $ and
     $ \lambda_{\gamma_1} = \lambda_{\gamma_2} = 0$,
 (c) $ \lambda_{e_1} = \lambda_{e_2} = 0.45 $ and
     $ \lambda_{\gamma_1} = \lambda_{\gamma_2} = 0$, and also
 (d) $ \lambda_{e_1} = \lambda_{e_2} = 0$ and
     $ \lambda_{\gamma_1} = \lambda_{\gamma_2} = 1.0$, for completeness.
The difference between $\lambda_{e}$'s being 0.45 and 0.5
is about $5\%$ for case (c) and less than $3\%$ for case (a)
in the invariant mass differential cross section.
At the NLC, the Higgs signal appears as a pronounced peak
in the $ZZ$ invariant mass distribution, up to $M_H = 390$ GeV.
We can find in Fig. 3 that
the ratio of Signal/Background is enhanced for,
$\lambda_{e_1} = \lambda_{e_2}$ close to $+0.5$
and $\lambda_{\gamma_1} = \lambda_{\gamma_2} =0$ in a broad range.

Figures 4 shows that the Higgs signal for $M_H = 400$ GeV is
visible in the invariant mass distribution of $ZZ$,
at $\sqrt{s_{e^+e^-}}$ = (a) 600, (b)700 and (c) 1000 GeV, in $\gamma\gamma$
collisions with photons from backscattered laser beams  for
$\lambda_{e_1} = \lambda_{e_2} = 0.45 $ and
$ \lambda_{\gamma_1} = \lambda_{\gamma_2} = 0 $.
The cross sections at the Higgs pole for
$\lambda_{e_1} = \lambda_{e_2} = 0 $ and
$ \lambda_{\gamma_1} = \lambda_{\gamma_2} = 0 $
are about $25\%$ smaller.
The combination of $\lambda_{e_1} = \lambda_{e_2} = 0 $ and
$\lambda_{\gamma_1} = \lambda_{\gamma_2} = \pm 1.0$ is slightly better
than unpolarized $e^+e^-$ and laser beams if $r < 0.5$ or $r > 0.8$.
A more realistic
study for the signal and background with the final states of
$l^+l^-\nu\bar{\nu}$ and $l^+l^-q\bar{q}$ is under investigation.

\section{Conclusions}
In high energy $\gamma\gamma$ collisions, the $TT$ cross section of
$\gamma\gamma \to ZZ$ dominates if $M_H >$ 350 GeV.
With $2\lambda_e \lambda_\gamma = 0$, the photon spectrum is almost flat,
and it is possible to search for the Higgs signal in a broad range below
$\sqrt{s_{e^+e^-}}$. The best case to search for the Higgs signal
is to have $\lambda_{e_1} = \lambda_{e_2}$ close to $+0.5$
and $\lambda_{\gamma_1} = \lambda_{\gamma_2} = 0$, because the
photon energy spectrum is almost flat and the contribution from
$\hat{\sigma}_{++LL}$ is enhanced by the Stokes parameters;
however, even in the best case, using polarized electrons or laser photons
yields only a small advantage over the totally unpolarized case.
At the NLC, a Higgs signal is possible for $M_H$ up to 390 GeV
in the invariant mass distribution of $ZZ$.
For larger CM energies, 600 GeV $< \sqrt{s_{e^+e^-}} <$ 1000 GeV,
it is possible to find a Higgs with a mass slightly larger than 400 GeV.
Thus there is not much advantage for this process in higher $e^+e^-$ energies.
Furthermore, there is not much advantage in the $\gamma\gamma$ mode;
$e^+e^-$ collisions at the NLC, by themselves, can search for the Higgs up to a
mass of 350 GeV.
The unique strength of high energy $\gamma\gamma$ collisions in the Higgs
search
is probably to measure the $H\gamma\gamma$ coupling with high precision
beyond the intermediate Higgs mass range.

\acknowledgements

The authors would like to thank David Bowser-Chao, Kingman Cheung
and George Jikia for discussions and Scott Willenbrock for suggestions
and comments.
This research was supported in part by DOE contracts DE-FG03-93-ER40757
and DE-FG05-87-ER40319.
%
%
\newpage
\appendix{RELEVANT FEYNMAN RULES IN THE NON-LINEAR GAUGE}

In this appendix, relevant Feynman rules
in the non-linear gauge described
in section 2 are presented in our conventions.
There are 3- and 4-point vertices among
the gauge boson fields, $W^{\pm}_\mu$, $Z_\mu$, $A_\mu$;
the Nambu-Goldstone boson fields, $G^{\pm}$, $G^0$;
the Fadeev-Popov ghost fields, $\theta^{\pm}$, $\bar{\theta}^{\pm}$,
$\theta^Z$, $\bar{\theta}^Z$, $\theta^A$, $\bar{\theta}^A$;
and the Higgs boson field $H$.
The gauge parameters are all taken to be unity,
$\xi_W = \xi_Z = \xi_A =1$,
which corresponds to a non-linear 't Hooft-Feynman gauge.
In this gauge, the $W$ boson ($W^{\pm}$), the Nambu-Goldstone boson
($G^{\pm}$),
and the Fadeev-Popov ghosts ($\theta^{\pm}$)
have the same mass $M_W = M_G =M_\theta$.
The Feynman rules for relevant interactions involved in the $W$, $G$
and $\theta$ loops which appear in the reaction
$\gamma\gamma \to ZZ$ are shown in Table IV.
%
%

%
%
%
%
%
\begin{table}
\caption{
The effect of $m_t$ on the cross section of $\gamma\gamma ZZ$
in $fb$ at $\sqrt{s_{\gamma\gamma}} = M_H = 300, 400, 500, 600, 700,$ and
800 GeV,  in the $++LL$ helicity state,
for $m_t = 120, 140, 160, 180$ and 200 GeV. }
\begin{tabular}{ccccccc}
$m_t$ (GeV)/ $M_H$ (GeV) & 300 & 400 & 500 & 600 & 700 & 800 \\
\tableline
120 & 360 & 55  & 22 & 14  & 11   &  9.4 \\
140 & 660 & 57  & 13 & 8.1 & 7.1  &  6.6 \\
160 & 790 & 87  & 11 & 4.2 & 4.0  &  4.2 \\
180 & 810 & 160 & 16 & 2.6 & 1.8  &  2.4 \\
200 & 830 & 210 & 29 & 3.3 & 0.69 &  1.1
\end{tabular}
\end{table}

\medskip

%
\begin{table}
\caption{
The total cross section of $\gamma\gamma ZZ$ in $fb$
at $\sqrt{s_{\gamma\gamma}} = M_H = 300, 400$ and 500 GeV,
in each helicity states
for $m_t = 140$ GeV after different cuts on the CM scattering angle
$|\cos(\theta^*)| < z_0$: $z_0 = 1.0, \cos( 30^o )$ and 0.8. }
\begin{tabular}{ccccccc}
$z_0$/ Helicities & $++LL$ & $++TT$ & $++LT$ & $+-LL$ & $+-TT$ & $+-LT$ \\
\tableline
(a) $\sqrt{s_{\gamma\gamma}} = M_H = 300$ GeV \\
1.0            & 660 & 160 & 0.099 & 1.4 & 47 & 8.0 \\
$\cos( 30^o )$ & 580 & 130 & 0.073 & 1.4 & 32 & 7.5 \\
0.8            & 530 & 120 & 0.057 & 1.3 & 26 & 7.1 \\
(b) $\sqrt{s_{\gamma\gamma}} = M_H = 400$ GeV \\
1.0            &  57 & 180 & 0.061 & 1.1 & 74 & 5.5 \\
$\cos( 30^o )$ &  49 & 150 & 0.037 & 1.1 & 45 & 5.1 \\
0.8            &  46 & 130 & 0.027 & 1.1 & 35 & 4.8 \\
(c) $\sqrt{s_{\gamma\gamma}} = M_H = 500$ GeV \\
1.0            &  13 & 240 & 0.034 & 0.96 & 98 & 3.9 \\
$\cos( 30^o )$ &  12 & 180 & 0.017 & 0.95 & 53 & 3.5 \\
0.8            &  11 & 160 & 0.012 & 0.93 & 40 & 3.2
\end{tabular}
\end{table}

\medskip

%
%
\begin{table}
\caption{
Total cross section of $\gamma\gamma \to ZZ$ in $fb$ as a function of
$\sqrt{s_{e^+e^-}}$ with backscattered laser photons,
for $|\cos(\theta^*)| < 0.8$, $m_t = 140$ GeV,
$M_H = 300, 400 $ GeV and $\infty$,
and five combinations of polarizations of initial $e^+e^-$ and laser beams
with $\lambda_{e_1} = \lambda_{e_2} = \lambda_e$,
and  $\lambda_{\gamma_1} = \lambda_{\gamma_2} = \lambda_{\gamma}$. }
\begin{tabular}{cccccccc}
$M_H$ (GeV) /$\sqrt{s_{e^+e^-}}$ (GeV)
& 240 & 300 & 400 & 500 & 600 & 700 & 1000 \\
\tableline
(a) $\lambda_e = 0.5$, $\lambda_{\gamma} =-1.0$ & & & & & & & \\
300 &     0.55 & 8.9 & 62 & 48 & 58 & 67 & 76 \\
400 &     0.47 & 7.7 & 30 & 54 & 59 & 66 & 75 \\
$\infty$& 0.45 & 7.4 & 28 & 46 & 58 & 65 & 74 \\
(b) $\lambda_e = 0.5$, $\lambda_{\gamma} =+1.0$ & & & & & & & \\
300 &     2.8 & 1.5 & 18  & 44 & 61 & 78 & 104 \\
400 &     2.7 & 1.3 & 10  & 27 & 47 & 63 & 96  \\
$\infty$& 2.6 & 1.3 & 9.8 & 25 & 42 & 58 & 94  \\
(c) $\lambda_e = 0$, $\lambda_{\gamma} = 0$ & & & & & & & \\
300 &     0.39 & 4.5 & 26 & 37 & 48 & 57 & 72 \\
400 &     0.38 & 4.3 & 16 & 30 & 42 & 52 & 70 \\
$\infty$& 0.37 & 4.2 & 15 & 28 & 40 & 49 & 69 \\
(d) $\lambda_e = 0.5$, $\lambda_{\gamma} = 0$ & & & & & & & \\
300 &     0.19 & 4.1 & 38 & 49 & 62 & 72 & 89 \\
400 &     0.17 & 3.6 & 18 & 39 & 54 & 67 & 87 \\
$\infty$& 0.16 & 3.5 & 17 & 35 & 51 & 64 & 86 \\
(e) $\lambda_e = 0$, $\lambda_{\gamma} =+1.0$ & & & & & & & \\
300 &     0.29 & 4.8 & 25 & 35 & 47 & 55 & 77 \\
400 &     0.27 & 4.4 & 16 & 30 & 40 & 51 & 71 \\
$\infty$& 0.26 & 4.4 & 15 & 27 & 38 & 48 & 69
\end{tabular}
\end{table}

\medskip

%
%
\begin{table}
\caption{
The Feynman rules for relevant interactions involved in the $W$, $G$
and $\theta$ loops which appear in the reaction $\gamma\gamma \to ZZ$
as modified  by the non-linear gauge condition which is described in Sec 2.
These can be compared to the unmodified rules given in Ref. \cite{Aoki}.
(Note that not all of the rules below are modified.)
All momenta ($e.$ $g.$, $k$, $p$ and $q$) and charges
are incoming to the vertices.
$g_{\mu\nu} \equiv diag(+,-,-,-)$ is the metric tensor,
$s_W$ is $\sin \theta_W$ and $c_W$ is $\cos \theta_W$. }
\begin{tabular}{llll}
3-point vertices & Feynman Rules & 4-point vertices & Feynman Rules \\
\tableline
$A_\mu(k) W^+_\nu(p) W^-_\rho(q)$ &
$-e [ g_{\mu\nu}(k-p-q)_\rho$      &
$A_\mu A_\nu W^+_\rho W^-_\sigma$ & $-2 e^2 g_{\mu\nu} g_{\rho\sigma}$ \\
    & $+g_{\nu\rho} (p-q)_\mu$ & & \\
    & $+g_{\rho\mu}(q-k+p)_\nu]$ & & \\
$A_\mu(k) G^+(p) G^-(q)$ & $+e (p-q)_\mu$ &
$A_\mu A_\nu G^+ G^-$    & $+2 e^2 g_{\mu\nu}$ \\
$A_\mu(k) \theta^+(p) \bar{\theta}^+(q)$ & $+e (p-q)_\mu$ &
$A_\mu A_\nu \theta^+ \bar{\theta}^+$    & $+2 e^2 g_{\mu\nu}$ \\
$A_\mu(k) \theta^-(p) \bar{\theta}^-(q)$ & $-e (p-q)_\mu$ &
$A_\mu A_\nu \theta^- \bar{\theta}^-$    & $+2 e^2 g_{\mu\nu}$ \\
$Z_\mu(k) W^+_\nu(p) W^-_\rho(q)$ &
$-g c_W \{ g_{\mu\nu} [(k-p)_\rho +\frac{s_W^2}{c_W^2}q_\rho]$ &
$Z_\mu Z_\nu W^+_\rho W^-_\sigma$ &
$- g^2 c_W^2 [ 2g_{\mu\nu} g_{\rho\sigma}$ \\
    & $+g_{\nu\rho} (p-q)_\mu$ & &
$-\frac{1-2s_W^2}{c_W^4}( g_{\mu\rho}g_{\nu\sigma}$ \\
    & $+g_{\rho\mu}[(q-k)_\nu-\frac{s_W^2}{c_W^2}p_\nu] \}$ & &
$+g_{\mu\sigma}g_{\nu\rho})]$ \\
$Z_\mu(k) G^+(p) G^-(q)$ &$+\frac{1}{2}g (\frac{1-2s_W^2}{c_W}) (p-q)_\mu$ &
$Z_\mu Z_\nu G^+ G^-$    &
$+\frac{1}{2}g^2\frac{(1-2s_W^2)^2}{c_W^2} g_{\mu\nu}$ \\
$Z_\mu(k) \theta^+(p) \bar{\theta}^+(q)$   &
$-g c_W (\frac{s_W^2}{c_W^2}p_\mu +q_\mu)$ &
$Z_\mu Z_\nu \theta^+ \bar{\theta}^+$      & $-2e^2 g_{\mu\nu}$ \\
$Z_\mu(k) \theta^-(p) \bar{\theta}^-(q)$   &
$+g c_W (\frac{s_W^2}{c_W^2}p_\mu +q_\mu)$ &
$Z_\mu Z_\nu \theta^- \bar{\theta}^-$      & $-2e^2 g_{\mu\nu}$ \\
%
%
$H Z_\mu Z_\nu$     & $+\frac{g}{c_W} M_Z g_{\mu\nu}$ &
$Z_\mu A_\nu W^+_\rho W^-_\sigma$ &
$-e g c_W [2g_{\mu\nu} g_{\rho\sigma}$ \\
$H W^+_\mu W^-_\nu$ & $+g M_W g_{\mu\nu}$ & &
$-\frac{1}{c_W^2}( g_{\mu\rho}g_{\nu\sigma} +g_{\mu\sigma}g_{\nu\rho} )]$ \\
$H G^+ G^-$ & $-\frac{1}{2} g \frac{M_H^2}{M_W}$ &
$Z_\mu A_\nu G^+ G^-$    & $+e g (\frac{1-2s_W^2}{c_W}) g_{\mu\nu}$ \\
$H \theta^{\pm} \bar{\theta}^{\pm}$ & $-\frac{1}{2} g M_W$ &
$Z_\mu A_\nu \theta^{\pm} \bar{\theta}^{\pm}$ &
$+e g (\frac{1-2s_W^2}{c_W}) g_{\mu\nu}$
\end{tabular}
\end{table}
%
%
\newpage
%
\figure{
The cross section of $\gamma\gamma \to ZZ$ as a function of
$\sqrt{s_{\gamma\gamma}}$ for the $LL$ (solid), $TT$ (dash-dotted)
and $LT$ (dashed) helicity states of $ZZ$
in (a) $++$ and (b) $+-$ helicity states of the photon with
$m_t =$ 140 GeV.
The ++LL cross section is evaluated with
$M_H$ = 300, 400, 500, 800 GeV and $\infty$. }
%
\figure{
The cross section of $\gamma\gamma \to ZZ$ as a function
of $\sqrt{s_{\gamma\gamma}}$ in the $++LL$ state,
for fermion loops alone (dotted),
the $W$ loop alone (dashed) and the sum of all loops (solid),
with $m_t$ = 120, 160 and 200 GeV and $M_H =$ 500 GeV.
The $W$ loop cross section has been evaluated with $m_t = 160$ GeV. }
%
\figure{
Invariant mass distribution of $\gamma\gamma \to ZZ$ in high energy
photon photon collisions from laser backscattered photons,
for the NLC energy, $\sqrt{s_{e^+e^-}} =$  500 GeV, $m_t =$ 140 GeV, and
$M_H$ = 250, 300, 350 and 390 GeV.
The polarizations of the initial $e^-e^+$ and laser beams
are taken to be
(a) $\lambda_{e_1} = \lambda_{e_2} = 0.45$ and
    $\lambda_{\gamma_1} = \lambda_{\gamma_2} = -1.0$,
(b) $\lambda_{e_1} = \lambda_{e_2} = 0$ and
    $\lambda_{\gamma_1} = \lambda_{\gamma_2} = 0$,
(c) $\lambda_{e_1} = \lambda_{e_2} = 0.45$ and
    $\lambda_{\gamma_1} = \lambda_{\gamma_2} = 0$
and (d) $\lambda_{e_1} = \lambda_{e_2} = 0$ and
    $\lambda_{\gamma_1} = \lambda_{\gamma_2} = 1.0$. }
%
\figure{
Invariant mass distribution of $\gamma\gamma \to ZZ$ in high energy
photon photon collisions from laser backscattered photons
with polarizations of the initial $e^-e^+$ and laser beams
being $\lambda_{e_1} = \lambda_{e_2} = 0.45$ and
      $\lambda_{\gamma_1} = \lambda_{\gamma_2} = 0$,
for $m_t =$ 140 GeV, $M_H$ = 250, 300, 350, 400, 450, and 500 GeV, and
(a) $\sqrt{s_{e^+e^-}} =$  600 GeV (without $M_H$ = 500 GeV),
(b) $\sqrt{s_{e^+e^-}} =$  700 GeV, and
(c) $\sqrt{s_{e^+e^-}} =$  1000 GeV. }
\eject
%
%

\begin{references}
\bibitem{Barger}
V.~Barger, K.~Cheung, B.~A.~Kniehl and R.~J.~N.~Phillips, Phys. Rev. {\bf D46}
(1992) 3725.
\bibitem{Jack}
J.~F.~Gunion, to appear in {\sl Proceedings of the International Workshop on
Physics and Experiments with Linear $e^+e^-$ Colliders}, Hawaii, USA (1993),
UCD-93-24, and references therein.
\bibitem{Ginzburg1}
I.~F.~Ginzburg, G.~L.~Kotkin, V.~G.~Serbo and V.~I.~Telnov,
Nucl. Instrum. Methods {\bf 205}, (1983) 47.
\bibitem{Ginzburg2}
I.~F.~Ginzburg, G.~L.~Kotkin, S.~L.~Panfil, V.~G.~Serbo and V.~I.~Telnov,
Nucl. Instrum. Methods {\bf 219}, (1984) 5.
\bibitem{Barklow}
T.~L.~Barklow, SLAC-PUB-3564 (1990), to appear in
{\sl The Proceedings of the 1990 DPF Summer Study on High Energy Physics},
Snowmass (1990).
\bibitem{Telnov}
V.~I.~Telnov, Nucl. Instrum. Methods {\bf A294}, (1990) 72.
\bibitem{BBC} D.~L.~Borden, D.~A.~Bauer, D.~O.~Caldwell, SLAC preprint
SLAC-PUB-5715 (1992).
\bibitem{Higgs}
J.~F.~Gunion and H.~E.~Haber, {\sl The Proceedings of the 1990 Summer Study
on High Energy Physics}, Snowmass (1990);
J.~F.~Gunion and H.~E.~Haber, UCD-92-22 (1992).
\bibitem{Haber}
H.~E.~Haber,
in {\sl Proceedings of the 1st International Workshop on Physics
and Experiments with Linear $e^+e^-$ Colliders},
Saariselk\"{a}, Finland, 1992, World Scientific Publishing, Singapore, (1992).
\bibitem{Boos}
E.~E.~Boos and G.~V.~Jikia, Phys. Lett. {\bf B275}, (1992) 164.
\bibitem{Gunion}
J.~F.~Gunion, UCD-93-8 (1993).
\bibitem{Cheung}
D.~Bowser-Chao and K.~Cheung, Phys.~ Rev.~ {\bf D48}, 89 (1993).
\bibitem{Jikia}
G.~V.~Jikia, Phys. Lett. {\bf B298}, (1993) 224;
G.~V.~Jikia, IHEP 93-37 (1993).
\bibitem{Berger}
M.~S.~Berger, MAD/PH/771 (1993).
\bibitem{Fujikawa}
K.~Fujikawa, Phys. ReV. {\bf D7} (1973) 393.
\bibitem{Bace}
M.~Bace and N.~D.~Hari Dass, Ann. Phys. {\bf 94} (1975) 349.
\bibitem{Gavela}
M.~Gavela, G.~Girardi, C~Malleville and P.~Sorba,
Nucl. Phys. {\bf B193} (1981) 257.
\bibitem{Desh}
N.~G.~Deshpande and M.~Nazerimonfared, Nucl. Phys. {\bf B213} (1983) 390.
\bibitem{Boudjema}
F.~Boudjema, Phys. Lett. {\bf B187} (1987) 362.
\bibitem{ggZZ}
D.~A.~Dicus, C.~Kao,and W.~W.~Repko, Phys. Rev. {\bf D36} 1570 (1987);
D.~A.~Dicus, Phys. Rev. {\bf D38} 394 (1988).
\bibitem{Glover}
E.~W.~N.~Glover and J.~J.~van~der~Bij, Phys. Lett. {\bf B219}, 488 (1989);
Nucl.~ Phys.~ {\bf B321}, 561 (1989).
\bibitem{LOOP} D.~Dicus and C.~Kao, LOOP, a FORTRAN program for doing
loop integrals of 1, 2, 3 and 4 point functions with momenta in the numerator,
unpublished, (1991).
\bibitem{Tini} G.~'t~Hooft and M.~Veltman, Nucl. Phys. {\bf B153}, 365 (1979).
G.~Passarino and M.~Veltman, Nucl. Phys. {\bf B160}, 151 (1979).
\bibitem{Aoki}
K.-I.~Aoki, et al., Prog. Theo. Phys. Suppl. No. 73 (1982).
%
\end{references}
\end{document}